\documentclass[sigconf,nonacm]{acmart}
\usepackage[most]{tcolorbox}

\AtBeginDocument{%
  \providecommand\BibTeX{{%
    \normalfont B\kern-0.5em{\scshape i\kern-0.25em b}\kern-0.8em\TeX}}}

\setcopyright{acmcopyright}
\copyrightyear{2018}
\acmYear{2018}
\acmDOI{10.1145/1122445.1122456}

\acmConference[Woodstock '18]{Woodstock '18: ACM Symposium on Neural
  Gaze Detection}{June 03--05, 2018}{Woodstock, NY}
\acmBooktitle{Woodstock '18: ACM Symposium on Neural Gaze Detection,
  June 03--05, 2018, Woodstock, NY}
\acmPrice{15.00}
\acmISBN{978-1-4503-XXXX-X/18/06}

\begin{document}

\title[TUG: Predicting Interpersonal Compatibility]{Tacit Understanding Game (TUG): Predicting Interpersonal Compatibility}

\author{Yueshen Li}
\email{yueshen7@illinois.edu}
\affiliation{%
  \institution{University of Illinois Urbana-Champaign}
}

\author{Krishnaveni Unnikrishnan}
\email{ku18@illinois.edu}
\affiliation{%
  \institution{University of Illinois Urbana-Champaign}
}
\author{Aadya Agrawal}
\email{aadyaa3@illinois.edu}
\affiliation{%
  \institution{University of Illinois Urbana-Champaign}
}

\renewcommand{\shortauthors}{Li et al.}

\begin{abstract}
Research on relationship quality often relies on lengthy questionnaires or invasive textual corpora, limiting ecological validity and user privacy. We ask whether a sequence of single‑word choices made in a playful setting can reveal personality and predict interpersonal compatibility. We introduce the Tacit Understanding Game (TUG), a two-player online word association game. We collect word choice traces, annotate a subset with psychological ground truth scales, and bootstrap a larger synthetic corpus via large language model simulation. TUG demonstrates that minimal, privacy preserving signals can support relationship matching, offering new design space for social platforms.
\end{abstract}

\ccsdesc[500]{Human-centered computing~Empirical studies in HCI}
\ccsdesc[300]{Human-centered computing~Collaborative interaction}
\ccsdesc[100]{Computing methodologies~Machine learning}
\ccsdesc[100]{Security and privacy~Human and societal aspects of security and privacy}

\keywords{interpersonal compatibility, privacy‑preserving inference, word association game, human–computer interaction, personality modeling, machine learning}

\maketitle

\section{Introduction}
Interpersonal compatibility is a cornerstone of social well‑being that influences everything from relationship satisfaction to workplace productivity. In HCI, measuring and facilitating social cohesion is traditionally addressed through survey instruments (e.g., URCS \cite{Dibble12})  or text mining approaches that repurpose large social media corpora \cite{Yarkoni10}. Both approaches clash with modern design values: surveys impose cognitive load and presume existing relationships, while unfettered text collection raises privacy and inclusivity concerns. As social networking platforms shift toward lightweight privacy-respecting interactions, the field needs new techniques that capture rich interpersonal signals without intrusive data requests. We need a technique that (1) works across \emph{any} relationship type, (2) predicts \emph{potential} compatibility (not just retrospective closeness), and (3) demands only minimal user input.

\smallskip
We position the Tacit Understanding Game (TUG) within a growing body of HCI work that uses playful crowdsourcing to harvest behavioural data at scale. Inspired by ESPGame \cite{vonAhn04}, Foldit \cite{Cooper10}, and GuesstheKarma \cite{glenski18}, TUG leverages intrinsic motivation - fun, competition, curiosity - to generate single-word choices that double as entertainment feedback for players and structured input for machine learning models. The game design aligns with the principles of Games with a Purpose (GWAP): every click fuels a dataset while rewarding the contributor in real time.

\smallskip
Each 10-round TUG session produces 30–50 distinct word selections per pair of players, producing a high-density relational trace at marginal cost. Our pilot deployment collected 16 real pairs without monetary incentives, demonstrating the feasibility of organic data acquisition. To mitigate the cold‑start problem, we generate 6000 additional rounds through LLM‑based simulation that samples from the empirical distribution of theme-word co-occurrences. In HCI research, synthetic augmentation is increasingly used to seed recommender systems and calibrate crowd models when real data is scarce. TUG embeds a user-friendly feedback prompt at the end of the session, asking players to rate the clarity of the UI, the flow of the game, and the perceived fairness. Qualitative comments feed an affinity‑diagram‑driven redesign cycle, while quantitative ratings inform an active‑learning schedule that prioritizes misclassified pairs for UI tweaks and model fine‑tuning.

\smallskip
\noindent\textbf{Research questions.}
\begin{enumerate}
  \item \textbf{Personality Reflection}: Can a person be represented by their sequence of single‑word choices? And if so, to what degree?
  \item \textbf{Compatibility Prediction}: Can such representations be combined to forecast pair compatibility?
\end{enumerate}

\noindent\textbf{Contributions.}  This work offers:  
(1) a \emph{game-based data-collection method} that preserves privacy while scaling to diverse relationships,  
(2) a \emph{hybrid real + synthetic dataset} for compatibility modelling, and  
(3) an \emph{interactive ML pipeline} that refines both UI and model through continuous user feedback.  
Together, these contributions advance HCI–AI research on privacy-respecting social sensing.


\section{Related Work}
\subsection{Gamified Crowdsourcing and Games With a Purpose}
Early work in human-computation games demonstrated the power of gamified crowdsourcing for data collection. A seminal example is the ESP Game, where pairs of online players guessed words to label images \cite{vonAhn04}. This approach turned tedious image annotation into a fun activity and effectively harnessed player input to produce valuable labels (e.g. for improving image searches). Von Ahn and Dabbish’s ESP Game became a model for “Games With a Purpose,” showing that gameplay can crowdsource large-scale data that algorithms alone struggle to generate \cite{vonAhn08}. Subsequent GWAPs extended this idea to other domains: for instance, Foldit transformed protein folding into a game and enabled tens of thousands of players to help solve biochemical puzzles \cite{Cooper10}, KissKissBan for semantic relations \cite{ho10}, and Guess-the-Karma for social-media preference prediction \cite{glenski18}. These successes underscore how well-designed games can motivate participation and yield high-quality datasets in vision, science, and beyond. The Tacit Understanding Game (TUG) builds on this tradition of gamified data collection, but explores a new domain: instead of collecting image labels or scientific solutions, TUG crowdsources data about interpersonal understanding. In other words, TUG is a GWAP centered on human relationships – it engages pairs of people in a game that covertly elicits signals of their compatibility, transforming an abstract social sensing task into an enjoyable experience. By leveraging play, TUG aims to gather subtle personal data (e.g. word associations, mutual perceptions) at scale without the friction of surveys, much like ESP Game and Foldit did for their domains. This shift poses new design questions about privacy and data sparsity, which we address through (i) minimal input design, (ii) LLM-based synthetic augmentation, and (iii) iterative UI feedback—capabilities not leveraged in earlier GWAPs. Thus, TUG not only inherits the motivational benefits of playful interaction but also pushes GWAP methodology toward sensitive social-computing problems.

\subsection{Personality, Language, and Interpersonal Compatibility}
Interpersonal compatibility has traditionally been studied through direct questionnaires and psychometric instruments. For example, relationship closeness is often measured by self-report scales like the Unidimensional Relationship Closeness Scale (URCS)\cite{Dibble12}, and individual personality traits by instruments such as the Big Five Inventory (e.g. the 10-item BFI-10 short form)\cite{Rammstedt07}. These established tools (URCS for closeness, BFI for personality) provide validated ground truth for how close or compatible two people feel, but they require participants to explicitly answer personal questions. Researchers have also developed computational methods to infer personality and compatibility implicitly from language and behavior. Linguistic Inquiry and Word Count (LIWC) is a widely-used text analysis tool that maps word usage to psychological categories \cite{Tausczik10}. Studies using LIWC show that language patterns (e.g. pronoun use, emotional words) correlate with personality traits and even relationship dynamics. For instance, Ireland et al. found that language style matching between partners’ function words can predict relationship stability – a hint that subtle verbal cues signal compatibility \cite{Ireland11}. On the personality side, Yarkoni (2010) analyzed blogging data and showed that word frequencies in one’s writing reliably reflect Big Five personality traits \cite{Yarkoni10}. Intriguingly, recent work suggests we can go to an even lower granularity: Sakamoto et al. (2021) demonstrated that even a single self-chosen word can be used to estimate a person’s Big Five personality profile with reasonable accuracy \cite{Sakamoto21}. Similarly, Kosinski et al. (2013) showed that sparse digital footprints (e.g. a handful of Facebook “Likes”) can predict personal attributes like personality and political views with surprising precision \cite{Kosinski13}. These advances imply that rich psychological signals exist in minimal data – a motivating insight for TUG. The TUG project draws on this idea by using a single-word clue from each player as input; by analyzing how two people’s words or choices relate, TUG attempts to infer their compatibility without a full questionnaire. This approach aligns with prior findings that brief textual or behavioral cues can reveal interpersonal traits, yet TUG packages it in a game format. In summary, TUG’s design bridges social psychology measures (like URCS, BFI) and computational inference (LIWC-style text analysis, minimal-input personality prediction), using gameplay to elicit the small but telling signals that indicate when two people “click.”

\subsection{Privacy-Preserving Data Collection and Synthetic Data}
Collecting data on people’s relationships and personalities raises privacy concerns, so researchers have explored methods to gather or utilize such data in privacy-preserving ways. One approach is to limit the data collected to only low-dimensional or abstract signals – for example, using a single word or a few game actions rather than detailed personal disclosures \cite{Sakamoto21}. TUG embodies this approach: because players only share intuitive clues or guesses in a game (and not sensitive facts about themselves), the data remain one step removed from raw personal information. This indirect approach offers a layer of privacy by design, as the game never asks participants to explicitly reveal private traits or opinions about each other. Another complementary approach in the community is the use of synthetic or simulated datasets in social computing. For instance, Park et al. (2022) introduce Social Simulacra, a technique to generate realistic synthetic social media interactions for prototyping community designs \cite{Park22}. By populating a system with AI-generated user data, one can evaluate social computing systems without risking real users’ privacy. Similarly, recent studies have used large language models to produce synthetic user responses for research surveys or UX studies \cite{Kosinski13}. The use of simulated data highlights a trend of reducing dependence on real personal data. We augment our modest in-the-wild corpus with 6000 LLM-simulated rounds that mirror observed theme–word co-occurrence patterns. This hybrid corpus balances privacy, scale, and ecological validity. 

\smallskip
TUG’s gameplay data could be seen as a form of “minimally identifiable” data – it’s real user input, but in a gamified, abstract format that blurs personal identifiers. In effect, TUG aims for privacy-preserving crowdsourcing: it crowdsources a model of interpersonal compatibility (by learning from many game rounds) while minimizing the exposure of any individual’s private information. This design is in line with the broader goal of balancing data-driven inference with user privacy. Compared to directly asking users to fill out personality inventories or share relationship details, TUG’s game-like interface may feel less invasive, encouraging participation and honest input. Finally, by accumulating a large volume of small, privacy-safe clues, the system can still learn population-level patterns – much like aggregated digital traces have been used to infer traits without exposing any single user. In summary, TUG’s method resonates with emerging practices in social computing that favor data abstraction and synthesis to safeguard privacy while still enabling rich inferential power.

\subsection{Interactive Feedback Loops and Human-in-the-Loop Refinement}
An important aspect of HCI systems like TUG is the inclusion of interactive feedback loops that improve the system through continued user interaction. Prior work on interactive machine learning and human-in-the-loop systems has shown that models can be iteratively refined by user input and feedback. In an influential overview, Amershi et al. (2014) argued that giving end-users the ability to train, adjust, and correct models leads to better performance and user experience, effectively empowering people to shape the intelligent systems they use \cite{Amershi14}. TUG applies this principle in a couple of ways. First, the game provides immediate feedback to players – for example, revealing a score or outcome that reflects how well their inputs (words or choices) matched, which in theory represents the system’s current estimate of their compatibility. This feedback to the users not only makes the game engaging, but also closes the loop: users can adjust their strategy or provide new inputs in subsequent rounds, indirectly training the underlying compatibility model. Second, the data collected from each game round serve as feedback to the developers/researchers and the model itself. As more people play, TUG’s predictive model can be updated on the backend using the growing dataset of word associations and compatibility outcomes. This is analogous to the model refinement loops seen in interactive recommender systems or adaptive user interfaces, where each interaction informs better performance over time. For instance, in an interactive learning-to-rank system, user clicks or corrections continually refine the ranking algorithm. Likewise, TUG’s design enables a form of cumulative learning: every guess and rating from players is a new datapoint to reduce the model’s uncertainty about what signals predict rapport. The inclusion of humans in this loop – both as players generating data and as recipients of the game’s feedback – echoes the “Power to the People” ethos highlighted by Amershi et al., in which end-users guide the system’s evolution \cite{Amershi14}. Moreover, this synergy between UI and model can drive design refinement: if certain word prompts or interface elements consistently confuse players or yield poor predictions, designers can tweak the UI, and the model can adapt accordingly. In summary, TUG’s iterative gameplay and updating model represent a co-evolving feedback cycle. This aspect of the project situates it in a lineage of HCI work that marries user experience design with machine learning, ensuring that the system improves not only its predictive accuracy but also its alignment with how users actually behave and perceive feedback.


\section{Methodology}

Our methodology adopts a user-first, privacy-preserving approach across data collection, modeling, and iteration. At every stage, we aimed to reduce cognitive load, protect user privacy, and make the experience engaging through gameplay. To simplify interaction, we restricted inputs to single words rather than open-ended text. This ensured lexical parity across players and made the task intuitive. The pipeline is structured around six components: word matrix design, game design, real-world data collection, feedback-driven design improvements, synthetic data generation, and model training. Together, these components support our goal of modeling tacit compatibility through interaction data.

\subsection{Word Matrix Design}
\label{sec:word-matrix}

The core interaction in TUG is structured around a 5×4 word matrix presented to both players in each round. To balance semantic richness with usability, we developed a curated vocabulary organized around broad, interpretable themes. We began by identifying 15 high-level themes such as “Music \& Sound,” “Philosophy \& Morality,” and “Humor \& Wit” which we hypothesized to reflect key dimensions of tacit compatibility. Instead of prompting an LLM to generate hundreds of words per theme directly (which often resulted in repetition or technical jargon), we broke each theme into finer-grained subcategories and used GPT-4 to generate approximately 100 words per subcategory. This allowed us to capture a wide semantic range while maintaining coherence and contextual relevance. A complete taxonomy of themes and subcategories is provided in Appendix~\ref{app:themes}.

\subsection{Game Design}

TUG (Tacit Understanding Game) is a synchronous, web-based game designed to elicit shared associations between two players through collaborative word selection. The gameplay serves both as an engaging social experience and as a crowdsourcing mechanism for collecting compatibility data. Its design draws on principles from HCI and game-based learning—focusing on fairness, user motivation, and interpretability. Snapshots of the game UI can be found in Appendix C.

\subsubsection*{Pairing and Session Flow}
Players access the TUG web app \footnote{The beta app is available at: https://tug-mvp.web.app/login and through the Github repository: https://github.com/listenha/tug-project.} and are paired in one of two ways:
\begin{itemize}
    \item \textbf{Queue-based matching:} Two users waiting for a new game at the same time are paired automatically.
    \item \textbf{Tag-based matching:} Known pairs can coordinate by entering a shared tag before starting the game, resulting in immediate pairing. 
\end{itemize}
\noindent Each session consists of 10 rounds. In every round, both players are shown:
\begin{itemize}
    \item A central \textbf{keyword} drawn from a predefined thematic category.
    \item A \textbf{5×4 word matrix} associated with that theme.
    \item A \textbf{selection quota}, randomly chosen between 3 and 5, indicating how many words each player must select.
\end{itemize}
\noindent To prevent players from developing position-based coordination strategies, the word matrix is randomly shuffled for each player.

\subsubsection*{Gameplay Mechanics}
Players are instructed to choose the quota-sized set of words they most associate with the keyword. After both players submit their selections, they are:
\begin{itemize}
    \item Presented with a \textbf{round score} based on how many word choices matched.
    \item Shown the \textbf{matched words} in that round.
    \item Given the opportunity to \textbf{share one of their unmatched words} with their partner to promote reflection or connection.
\end{itemize}
\noindent The next round then begins. After 10 rounds, players return to the lobby and may start a new session.

\subsubsection*{Scoring System}
To fairly compare rounds with different selection quotas, scoring is based on the \textbf{Word Choice Match Rate (WCMR)}:
\[
\text{WCMR} = \frac{\text{Number of Matched Words}}{\text{Quota}}
\]
\noindent Points are scaled using a \textbf{round factor} that increases with quota size to reflect the greater difficulty of achieving high alignment:
\begin{center}
Quota 3 → 30,\quad Quota 4 → 40,\quad Quota 5 → 50,\quad etc.
\end{center}
\noindent Scores are rounded to the nearest 10 for simplicity. For instance, a 50\% WCMR with Quota 4 yields a score of 20.
\noindent \textbf{Consistency bonuses} are awarded to encourage sustained compatibility:
\begin{itemize}
    \item 3 consecutive rounds with WCMR $\geq$ 60\%: +50 points
    \item 5 consecutive rounds with WCMR $\geq$ 60\%: +150 points
\end{itemize}
\noindent Bonuses are cumulative, supporting streak-based engagement.

\subsubsection*{Leaderboard and Motivation}
A live leaderboard is displayed on the TUG landing page. It ranks players by total session score (across 10 rounds), introducing both intrinsic (self-awareness, shared cognition) and extrinsic (ranking, recognition) motivations. During the beta phase, the top 5 players on the leaderboard were awarded gift cards to further incentivize participation. Only completed 10-round sessions are eligible for the leaderboard, ensuring data quality and full-session context for compatibility modeling..

\subsection{Crowdsourcing Dataset}

To bootstrap real-world evaluation of TUG, we recruited 15 pairs of beta testers nominated by team members. Of these, 10 were couples in a relationship and 5 were friends. These early users were instrumental in generating high-quality, labeled gameplay data while helping us evaluate interface usability, semantic alignment and our ML model in a realistic setting.

\subsubsection*{Consent and Privacy Protocol}

In alignment with data ethics best practices outlined in the \textit{Datasheets for Datasets} framework \cite{gebru2021datasheetsdatasets}, we prioritized transparency and informed participation throughout our data collection process. Each participant received a detailed beta testing guide clearly outlining what data would be collected and how it would be used. Participants were explicitly informed that:
\begin{itemize}
    \item No personally identifiable information (PII) would be recorded.
    \item All gameplay logs would be anonymized and securely stored.
    \item Data would be used solely for research and iterative design improvement.
    \item They could withdraw at any time without penalty.
\end{itemize}

\noindent We believe that this emphasis on consent and clarity ensures that contributors understood the scope and purpose of their participation, and treats them as informed participants rather than passive data points.

\subsubsection*{Post-Game Feedback}

After completing a session, participants filled a brief optional feedback form to evaluate UI clarity, gameplay fairness, and flow on a 5-point Likert scale. Optional qualitative comments were also collected, offering rich insight into user experience.

\subsubsection*{Questionnaire-Guided Compatibility Scoring}

Rather than asking participants to self-rate their compatibility directly which can be biased or awkward, we adopted a psychology-informed approach. After each session, players completed two short questionnaires:

\paragraph{(1) Pair Questionnaire.}
The Unidimensional Relationship Closeness Scale (URCS) \cite{urcs2011close} was used to infer a compatibility score based on perceived closeness. The questionnaire includes 12 items rated on a 7-point IOS scale \cite{aron1992inclusion} (1 = Strongly Disagree, 7 = Strongly Agree) . Items capture aspects such as emotional connection, time spent together, and relational prioritization. While URCS is validated for various relationship types, some friend pairs reported discomfort when responding to certain intimacy-framed items. This highlighted a key limitation of explicit questionnaires and further motivated our focus on implicit, gameplay-based compatibility scoring.

\paragraph{(2) Individual Questionnaire.}
Each player also completed a 10-item personality self-assessment adapted from the Big Five Inventory (BFI10) \cite{rammstedt2007bfiten}, measuring traits such as extraversion, openness, and conscientiousness. These individual traits were collected for use in downstream analysis and model evaluation but were not used to compute compatibility scores directly in this study phase.
\\\\
Together, the pair and individual questionnaires provided a soft ground truth for evaluating compatibility predictions, while also informing future model calibration and generalization.
\subsection{Iterative Design Improvements}
\label{sec:design-improvements}

TUG was developed using a participatory, feedback-driven design process. Input from our initial beta testers directly informed improvements to accessibility, interpretability, and overall player experience. Our goal throughout was to reduce friction, minimize cognitive load, and ensure that TUG remains intuitive and enjoyable.

\subsubsection*{Difficulty Filtering} One of the most consistent points of feedback was that many words in the original word matrix felt too obscure or academic similar to "GRE-level" vocabulary. This was especially challenging for non-native English speakers and often disrupted the intended flow of play. To address this, we implemented a two-step word familiarity filtering pipeline. First, we manually rated a seed set of words per theme on a 1–6 difficulty scale (1 = very familiar, 6 = rare or domain-specific). Using this as a few-shot prompt, we then asked Google Gemini to rate the remaining words. Only words rated 1–3 were retained, resulting in a curated set of approximately 250 accessible words per theme. This reduced player confusion and cognitive fatigue while preserving semantic richness.

\subsubsection*{Chinese Word Matrix Development} In parallel with the English gameplay experience, we are developing a standalone Chinese version of the TUG word matrix. Rather than directly translating the English word lists, we are curating a native Chinese matrix through an equivalent design pipeline as described in Section ~\ref{sec:word-matrix} tailored to Chinese linguistic and cultural contexts. This decision emerged from our observation that many contextual and cultural cues were lost when attempting to directly translate the English word matrix into Chinese. Words that carry metaphorical or emotional weight in English often fail to evoke the same associations in Chinese, which undermines the tacit, intuitive gameplay that TUG aims to capture.

\subsubsection*{Contextual Word Definitions} We introduced lightweight semantic scaffolding in the form of contextual word definitions. These definitions were fetched using the wikipedia Python package and displayed on-demand when users hovered over a small info icon next to each word. This design allowed users to quickly access clarifications without leaving the game interface. The inclusion of definitions supported on-the-fly learning and ensured that even moderately unfamiliar words remained approachable, preserving engagement while reducing cognitive load.

\subsubsection*{Overall User Feedback and Reflections} Players consistently described TUG as 'fun','surprisingly deep', and 'insightful.' Many reported uncovering unexpected shared interests or recalling personal memories tied to specific word associations. These reactions align with our core design goal: to create a socially meaningful, cognitively light experience that surfaces tacit compatibility through gameplay. 


\subsection{Synthetic Data Generation}
To address the lack of large-scale real gameplay data, we adopted a human-centered, generative simulation approach that integrates language models as surrogate evaluators of semantic alignment. Our pipeline consists of three stages: round simulation, LLM-based scoring, and session assembly.
\\
\noindent Before settling on this pipeline, we experimented with prompting large language models to directly generate word choices for player pairs conditioned on a target compatibility score. However, this approach tended to produce overly deterministic or biased outputs, often failing to reflect the unpredictability and variance observed in real gameplay. As a result, we opted for a more bottom-up strategy: simulating realistic rounds first using theme-aware sampling and then using LLMs to evaluate compatibility post hoc. This structure better preserved natural randomness while enabling scalable, soft-labeled supervision.

\subsubsection*{Round-Level Simulation} To bootstrap training in the absence of large-scale real-world interaction data, we generated synthetic gameplay rounds using theme-specific word pools curated via Sentence-BERT embeddings \cite{reimers2019sentencebertsentenceembeddingsusing}. For each synthetic round, we first sampled 21 words from a randomly selected theme. We computed Sentence-BERT embeddings for all 21 words and identified the centroid in embedding space. The word closest to this semantic centroid was designated as the keyword, while the remaining 20 words formed the word matrix shown to both players. A selection quota between 3 and 5 was then chosen at random. We simulated two independent players, each selecting words they “associate most strongly” with the keyword—modeled as individual agents drawing from the same matrix. This process yielded rounds that reflect divergent yet plausible interpretations of a shared prompt, mimicking the variability observed in real player behavior.

\subsubsection*{LLM-Based Round Scoring} To evaluate conceptual alignment between simulated players, we leveraged Google Gemini 2.0 Flash as a semantic scoring oracle. Using a few-shot prompting setup (inspired by methods in \cite{wu2023evaluating} and \cite{subramonyam2023whyjohnny}), we framed the task as a reflective judgment of shared understanding. Prompts included exemplar rounds with annotated scores to calibrate the model’s responses, prioritizing meaning over surface overlap. This design acknowledges the challenges non-experts face in crafting effective LLM prompts, as outlined by Subramonyam et al. \cite{subramonyam2023whyjohnny}, and illustrates how structured scaffolding can guide LLM reasoning toward HCI-relevant criteria like mutual intelligibility. \footnote{See Appendix~\ref{app:prompt} for the full prompt template used during LLM-based scoring.}

\subsubsection*{Pairwise Session Assembly} Scored rounds were grouped into fixed-length sessions to simulate extended player interactions. To ensure coverage across the compatibility spectrum, we stratified rounds into buckets based on LLM-assigned scores (e.g., [0.0–0.2], [0.8–1.0]) and sampled 10-round sessions with target distributions. Each session included synthetic player IDs, full interaction logs, and an aggregated compatibility score (mean of round scores). This process approximates naturalistic variance in gameplay behavior and enables training on a surrogate dataset where compatibility is operationalized through LLM-based semantic judgments.
\\\\Our pipeline reflects an emerging paradigm in HCI research where synthetic data generation is not merely a fallback but a form of design intervention—co-creating interactional datasets with LLMs as judgment agents. Echoing findings from Wu et al. \cite{wu2023evaluating}, we emphasize the importance of prompt clarity, example calibration, and score consistency to preserve construct validity in LLM-labeled data. Rather than relying on raw generative output, our approach embeds structure and controlled variation, foregrounding intentionality in both simulation and supervision. Using this framework, we generated a synthetic dataset comprising 400 player pairs, each with 10 rounds of gameplay—resulting in 4,000 scored round-level logs suitable for training and validation of our ML pipeline.

\subsection{Model Architecture}
\begin{figure*}[!h]
    \centering
    \includegraphics[width=\textwidth]{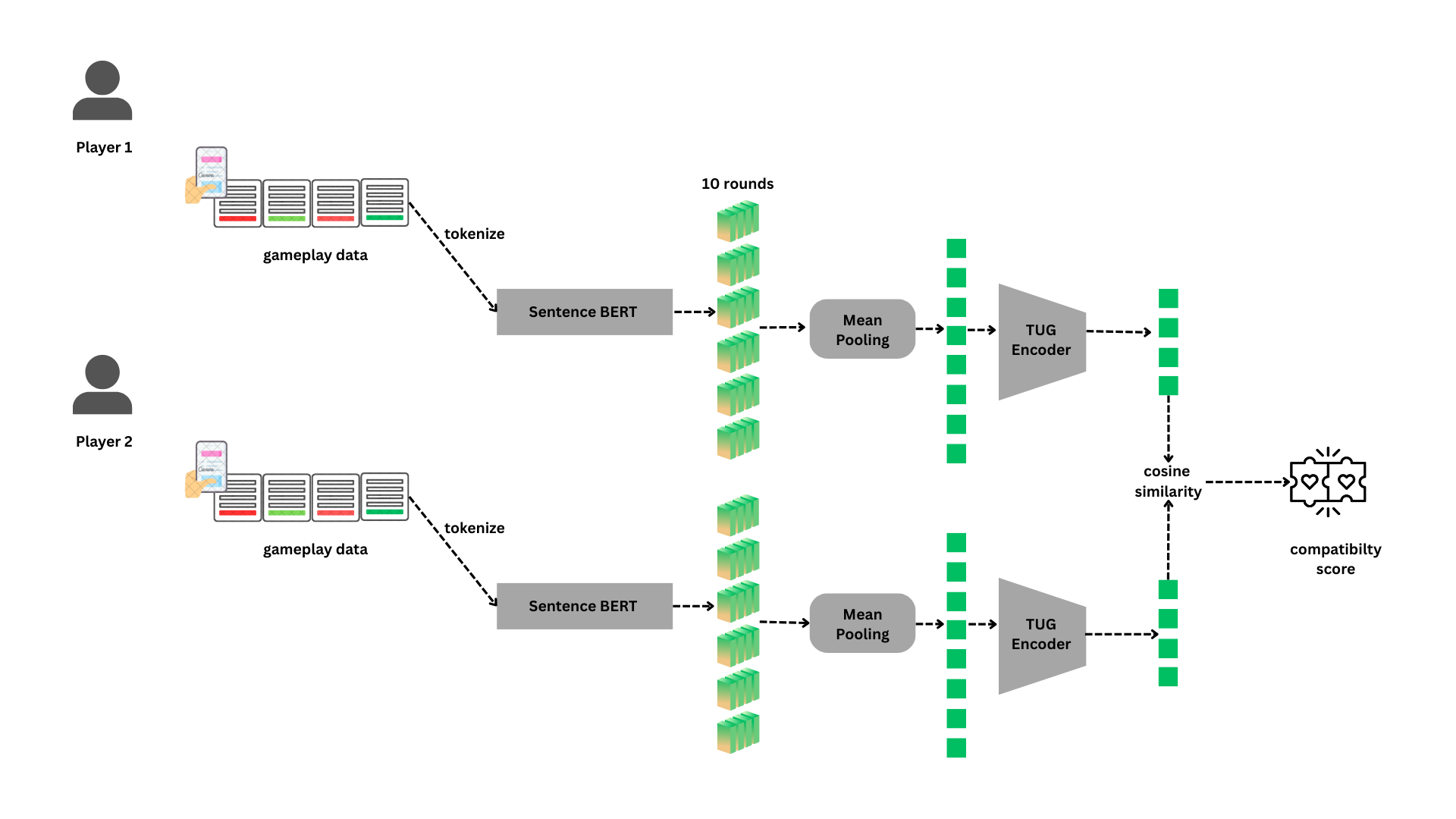}
    \caption{Siamese architecture for compatibility prediction from gameplay logs.}
    \label{fig:model-arch}
\end{figure*}

We model compatibility prediction using a Siamese neural network architecture that encodes gameplay data into latent vectors and compares them via cosine similarity. An overview is shown in Figure~\ref{fig:model-arch}.

Each player’s gameplay—10 rounds of selected words—is converted into round-level text (theme, keyword, selected words) and embedded using Sentence-BERT (SBERT), yielding 384-dimensional vectors per round. We apply mean pooling across rounds to obtain a single 384-dimensional embedding per player.

These embeddings are passed through a shared MLP encoder with two layers (384$\rightarrow$256$\rightarrow$128), ReLU activations, and layer normalization, producing 128-dimensional latent vectors $z_1$ and $z_2$. Compatibility is predicted via cosine similarity:

\[
\hat{y} = \frac{z_1 \cdot z_2}{\|z_1\| \|z_2\|}
\]

\subsubsection*{Auxiliary Supervision}

To regularize training, each player’s latent vector is also passed through a small regression head predicting $\hat{y}_1$ and $\hat{y}_2$, individual estimates of compatibility.

\subsubsection*{Training}

The model is trained on synthetic pairs with known scores $y$ (LLM-generated) using the loss:

\[ 
\mathcal{L} = \text{MSE}\left(\hat{y}, y\right) + \alpha \left[ \text{MSE}\left(\hat{y}_1, y\right) + \text{MSE}\left(\hat{y}_2, y\right) \right]
\]

with $\alpha=0.1$. We use Adam (lr = $10^{-3}$), early stopping on validation loss, and an 80/20 train/val split. Real human-scored data is reserved for out-of-sample evaluation only.

\section{Evaluation}

\subsection{Metrics}

We evaluated the model on a real gameplay dataset labeled with human-assigned compatibility scores in $[0,1]$. As the task is inherently continuous, we treat it primarily as a regression problem. We report:

\begin{itemize}
    \item \textbf{Pearson correlation}: Measures rank-order alignment between predicted and ground-truth scores.
    \item \textbf{Mean Absolute Error (MAE)} and \textbf{Root Mean Squared Error (RMSE)}: Quantify average prediction error magnitudes.
\end{itemize}

We also report classification metrics by thresholding scores (e.g., threshold = 0.75) to label pairs as “compatible” or “not compatible,” and compute accuracy, precision, recall, and F1-score. This reflects potential real-world use cases where binary matchmaking decisions are needed.


\subsection{Results}

Table~\ref{tab:evaluation} summarizes the model's performance on 15 real player pairs. The model achieved a Pearson correlation of 0.58, indicating moderate agreement with human rankings. MAE was 0.085 and RMSE 0.123, showing that predictions were typically within 9 percentage points of the true score.

\begin{table}[h]
    \centering
    \caption{Evaluation metrics on real gameplay data (15 pairs). Classification threshold: 0.75}
    \label{tab:evaluation}
    \begin{tabular}{l c}
        \toprule
        \textbf{Metric} & \textbf{Score} \\
        \midrule
        Pearson Correlation & 0.58 \\
        MAE & 0.085 \\
        RMSE & 0.123 \\
        Accuracy (binary) & 73.3\% \\
        Precision & 76.9\% \\
        Recall & 90.9\% \\
        F1-score & 83.3\% \\
        \bottomrule
    \end{tabular}
\end{table}

The classification-style evaluation yielded 73\% accuracy and strong recall (91\%), suggesting that the model effectively identifies most highly compatible pairs, albeit with a few false positives. The balanced F1-score (83\%) confirms robust binary performance despite the small dataset.

\subsection{Analysis}

\subsubsection*{Performance on Real Data}
Despite being trained solely on synthetic data, the model generalizes reasonably to human-labeled pairs. The moderate correlation and low MAE indicate that it captures semantic patterns aligned with human judgment. For example, if the true compatibility was 0.80, predictions typically ranged between 0.72 and 0.85.

A minor upward bias in predictions was observed—likely due to optimistic LLM-generated training labels or limited representation of low-compatibility pairs. This could be mitigated by collecting more diverse human gameplay logs.

Our evaluation confirms that the model captures key compatibility signals from gameplay choices. While results are promising, further improvements could come from real-data fine-tuning or domain adaptation. This approach shows potential for enhancing matchmaking and collaborative systems using implicit behavioral cues.

\subsubsection*{Precision–recall Trade-off}
The high recall (90.9 \%) at a 0.75 threshold indicates that the model rarely misses genuinely compatible pairs, a desirable property for matchmaking.  
However, precision (76.9 \%) lags, suggesting occasional false positives.  
A stricter threshold (0.80) lifts precision to 82.6 \% but drops recall to 84.8 \%, enabling practitioners to tune the operating point for their risk profile.





\subsection{Answering the Research Questions}

\begin{enumerate}
  \item \textit{RQ1 (Personality Reflection).}  
        Significant correlations between latent dimensions and Big-Five traits, together with a global $r=.58$ against URCS, show that single-word traces encode psychologically meaningful information.
  \item \textit{RQ2 (Compatibility Prediction).}  
        The recall value of 90.9\% affirms that learned representations capture compatibility beyond surface overlap.
\end{enumerate}
 

\section{Discussion}
This study presents the Tacit Understanding Game (TUG), a novel approach combining gamified crowdsourcing, minimal data collection, and synthetic data augmentation to predict interpersonal compatibility. By demonstrating that single-word choices made during a short, playful interaction can accurately reflect underlying personality traits and relationship closeness, our work significantly broadens the scope of how compatibility can be inferred in privacy-sensitive contexts. This is particularly impactful in a time when users increasingly demand transparency and minimalism in data collection practices. By validating this approach through rigorous experimentation, we position TUG as a viable alternative to traditional methods that rely on extensive self-disclosure or invasive text-mining of social media data.

\smallskip
One important implication of our results is the potential for TUG to integrate seamlessly into various social platforms and applications. For instance, social networking and dating apps could adopt TUG's playful, lightweight mechanism to facilitate matching processes, allowing users to obtain compatibility assessments without explicitly sharing sensitive personal details. Similarly, workplace and team-building applications could leverage TUG to assess team dynamics quickly and unobtrusively, potentially enhancing collaboration and productivity. Moreover, educational platforms and co-living communities could incorporate TUG as part of onboarding processes, helping new users form meaningful connections in a safe, privacy-respecting environment.

\subsubsection*{Limitations} Despite these promising results, our work has several limitations worth noting. The relatively small sample size of 15 real pairs restricts the generalizability of our findings, suggesting a need for larger-scale deployments to ensure robustness across diverse populations and relationship types. Moreover, reliance on synthetic data augmentation—though effective for initial model training—introduces potential biases from the underlying large language models, which might propagate cultural or linguistic stereotypes that merit careful examination in future studies. Additionally, our current modeling approach does not adequately capture pragmatic elements such as sarcasm or polysemous language, leading to potential mispredictions and highlighting the need for advanced language modeling techniques that handle these nuanced linguistic phenomena.

\subsubsection*{Plan of Work} Looking ahead, several avenues for future research emerge from this initial investigation. First, significantly expanding the real-world dataset through broader recruitment efforts would enhance the reliability of the model and allow nuanced analyses of demographic or cultural differences in interpersonal compatibility perception. Second, we aim to extend TUG by incorporating temporal modeling techniques, such as Transformer-based encoders, to capture evolving dynamics within interactions across game rounds. Third, conducting thorough fairness and bias audits across diverse user groups will be crucial to ensure equitable outcomes in real-world deployments. Finally, exploring explainability mechanisms, such as SHAP or LIME, to provide transparent feedback on compatibility predictions would further empower users by offering clear insights into the factors influencing their assessments.

\smallskip
Overall, our study establishes TUG not only as a methodologically innovative contribution to the intersection of HCI and computational social science but also as a practical framework that aligns with contemporary standards for ethical data use. By demonstrating how minimal, playful interactions can serve as proxies for richer psychological and relational data, we provide researchers and designers with a valuable template for future developments in privacy-conscious social computing.

\appendix
\section*{Appendix}
\section{Themes and Subcategories Used in TUG}
\label{app:themes}

Each round in TUG is built around a central keyword and a set of 20 associated words drawn from a specific theme. The following table outlines the 15 high-level themes we curated, each designed to reflect diverse dimensions of personal identity, preference, and cognitive style. Words were generated using LLMs, organized into subcategories, and filtered by difficulty and cultural accessibility (see Section \ref{sec:word-matrix} for details).

\begin{itemize}
    \item \textbf{Adventure \& Exploration:}
        \begin{itemize}
            \item Settings \& Activities: jungle, mountain, island, hiking, diving, backpacking, rafting
            \item Objects \& Tools: compass, map, lantern, rope, boots, tent, gear
            \item Feelings \& Motivations: thrill, wanderlust, courage, spontaneity, curiosity, freedom
            \item Roles \& Archetypes: explorer, adventurer, nomad, seeker, pathfinder
        \end{itemize}
    
    \item \textbf{Creativity \& Imagination:}
        \begin{itemize}
            \item Domains: art, fiction, fantasy, storytelling, design
            \item Processes: sketching, daydreaming, brainstorming, inventing
            \item Tools: brush, canvas, pen, clay, color, metaphor
            \item Archetypes: artist, dreamer, visionary, maker
        \end{itemize}

    \item \textbf{Science \& Logic:}
        \begin{itemize}
            \item Abstract Concepts: curiosity, skepticism, rigor, deduction
            \item Disciplines: physics, math, neuroscience, computer science
            \item Methods: hypothesis, modeling, theorizing, analysis
            \item Tools: microscope, theorem, algorithm, circuit
            \item Archetypes: scientist, philosopher, skeptic, analyst
        \end{itemize}

    \item \textbf{Philosophy \& Morality:}
        \begin{itemize}
            \item Virtues: justice, compassion, courage, humility
            \item Concepts: freewill, identity, paradox, causality
            \item Dilemmas: betrayal, fairness, intention, responsibility
            \item Schools: stoicism, nihilism, utilitarianism, deontology
            \item Archetypes: sage, rebel, monk, moralist
        \end{itemize}

    \item \textbf{Music \& Sound:}
        \begin{itemize}
            \item Genres: jazz, lo-fi, blues, techno, classical
            \item Instruments: guitar, violin, synthesizer, drum
            \item Roles: DJ, composer, listener, performer
            \item Actions: mixing, freestyling, composing, harmonizing
            \item Moods: nostalgic, upbeat, melancholic, rebellious
        \end{itemize}

    \item \textbf{Movies \& Pop Culture:}
        \begin{itemize}
            \item Genres: thriller, romcom, dystopia, superhero
            \item Characters: antihero, diva, mentor, sidekick
            \item Icons: Tarantino, Gaga, Zendaya, Spielberg
            \item Symbols: lightsaber, crown, cape, wand
            \item Fan Culture: fandom, meme, premiere, rewatch
        \end{itemize}

    \item \textbf{Literature \& Storytelling:}
        \begin{itemize}
            \item Genres: satire, fable, myth, tragedy
            \item Themes: identity, loss, justice, redemption
            \item Characters: orphan, sage, trickster, narrator
            \item Authors: Orwell, Brontë, Tolstoy, Dickinson
            \item Reader Types: annotator, dreamer, skimmer, critic
        \end{itemize}

    \item \textbf{Sports \& Competition:}
        \begin{itemize}
            \item Sports: soccer, fencing, tennis, boxing
            \item Roles: goalie, captain, umpire, coach
            \item Icons: Serena, Messi, Jordan, Nadal
            \item Objects: ball, net, whistle, helmet
            \item Values: rivalry, passion, teamwork, resilience
        \end{itemize}

    \item \textbf{Food \& Culinary:}
        \begin{itemize}
            \item Ingredients: cheese, tofu, basil, avocado
            \item Dishes: ramen, curry, sushi, biryani
            \item Tastes: spicy, creamy, chewy, umami
            \item Tools: blender, skillet, whisk, knife
            \item Identities: vegan, snacker, chef, gourmet
        \end{itemize}

    \item \textbf{Humor \& Wit:}
        \begin{itemize}
            \item Styles: sarcasm, satire, parody, slapstick
            \item Archetypes: jester, troll, fool, heckler
            \item Reactions: witty, cringy, savage, goofy
            \item Formats: meme, skit, roast, improv
            \item Humor Roles: classclown, bantermaster, observer, punner
        \end{itemize}

    \item \textbf{Technology \& Innovation:}
        \begin{itemize}
            \item Tools: app, sensor, cloud, database
            \item Fields: AI, fintech, cleantech, robotics
            \item Roles: coder, innovator, futurist, UXer
            \item Mindsets: scalability, disruption, iteration, creativity
            \item Trends: Web3, AGI, automation, smartcity
        \end{itemize}

    \item \textbf{Nature \& Outdoors:}
        \begin{itemize}
            \item Environments: forest, ocean, desert, meadow
            \item Activities: hiking, stargazing, camping, birdwatching
            \item Elements: tree, river, mountain, sky
            \item Animals: deer, fox, eagle, frog, bear
            \item Vibes: calm, wild, refreshing, grounded
        \end{itemize}

    \item \textbf{Social Bonds \& Relationships:}
        \begin{itemize}
            \item Traits: empathy, trust, closeness, loyalty
            \item Relationship Types: friendship, partnership, mentorship
            \item Dynamics: introvert, extrovert, communicator, listener
            \item Values: support, independence, belonging, intimacy
            \item Roles: confidant, bestie, partner, ally
        \end{itemize}

    \item \textbf{Work \& Productivity:}
        \begin{itemize}
            \item Styles: structured, flexible, multitasker, focused
            \item Tools: planner, calendar, laptop, to-do list
            \item Values: ambition, discipline, balance, hustle
            \item States: burnout, flow, distraction, grind
            \item Roles: leader, collaborator, soloist, brainstormer
        \end{itemize}

    \item \textbf{Dreams \& Aspirations:}
        \begin{itemize}
            \item Goals: impact, freedom, legacy, mastery
            \item Themes: self-discovery, purpose, identity, resilience
            \item Directions: realism, idealism, ambition, introspection
            \item Metaphors: path, ladder, compass, wings
            \item Archetypes: dreamer, doer, seeker, visionary
        \end{itemize}
\end{itemize}

\section{LLM Prompt Design for Compatibility Scoring}\label{app:prompt}

To generate round-level compatibility scores, we designed a structured few-shot prompt for Gemini 2.0 that emphasizes conceptual and semantic alignment between player responses. The prompt was iteratively refined to ensure interpretability, consistency, and alignment with our scoring goals. Inspired by HCI prompt design studies \cite{wu2023evaluating,subramonyam2023whyjohnny}, we explicitly framed the task using illustrative examples, value boundaries (0.0 to 1.0), and instructions to avoid lexical bias.

\subsection*{Prompt Template (Few-Shot)}

\begin{tcolorbox}[
    colback=gray!5!white,
    colframe=black!50,
    title=LLM Prompt for Compatibility Scoring,
    fonttitle=\bfseries,
    sharp corners,
    boxrule=0.5pt,
    width=\linewidth
]
Two players were shown the same keyword and asked to select a few words that they associate most strongly with that keyword. Based on how similarly they interpret and relate to the keyword, rate their compatibility on a scale from 0.0 to 1.0, where:

\begin{itemize}
    \item 1.0 = Their choices reflect very similar thinking or emotional framing — as if they share a mental model or perspective.
    \item 0.0 = Their choices suggest very different associations or understandings — little overlap in thought or tone.
    \item Values in between reflect partial similarity or moderate overlap in their thinking.
\end{itemize}

Avoid over-rewarding exact word matches. Instead, consider the underlying concepts, emotional tone, or shared associations reflected in the chosen words.\\
\textbf{Example 1} \\
Keyword: \texttt{"adventure"} \\
Player 1 Choices: \texttt{["explore", "voyage", "map", "treasure"]} \\
Player 2 Choices: \texttt{["sail", "pirate", "treasure", "ruins"]} \\
Score: \texttt{0.82}\\
Why: Both players imagine a classic treasure hunt scenario — shared metaphor, even with different words.\\
\textbf{Example 2} \\
Keyword: \texttt{"freedom"} \\
Player 1 Choices: \texttt{["liberty", "wilderness", "unbound"]} \\
Player 2 Choices: \texttt{["justice", "equality", "dignity"]} \\
Score: \texttt{0.55}
\\
Why: Both relate to freedom but from different angles — one personal/spatial, the other political/social.\\
\textbf{Example 3} \\
Keyword: \texttt{"technology"} \\
Player 1 Choices: \texttt{["robot", "AI", "automation"]} \\
Player 2 Choices: \texttt{["nature", "soul", "intuition"]} \\
Score: \texttt{0.10}\\
Why: These choices reflect very different worldviews — one mechanistic, the other humanistic.\\
\bigskip
Now evaluate the following:
\end{tcolorbox}

\noindent At inference time, this header was followed by a theme, keyword, and two players’ choices, prompting the LLM to respond with a floating-point score in $[0.0, 1.0]$.

\section{User Interface Snapshots}

\begin{figure}[H]
  \centering
  \includegraphics[width=0.8\linewidth]{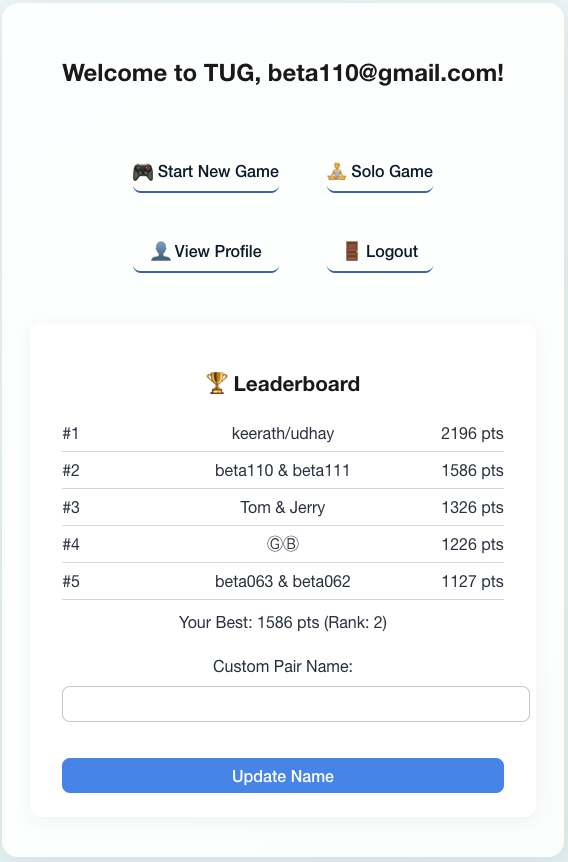}
  \caption{Game's home page and leaderboard.}
  \label{fig:lobby}
\end{figure}

\begin{figure}[H]
  \centering
  \includegraphics[width=0.8\linewidth]{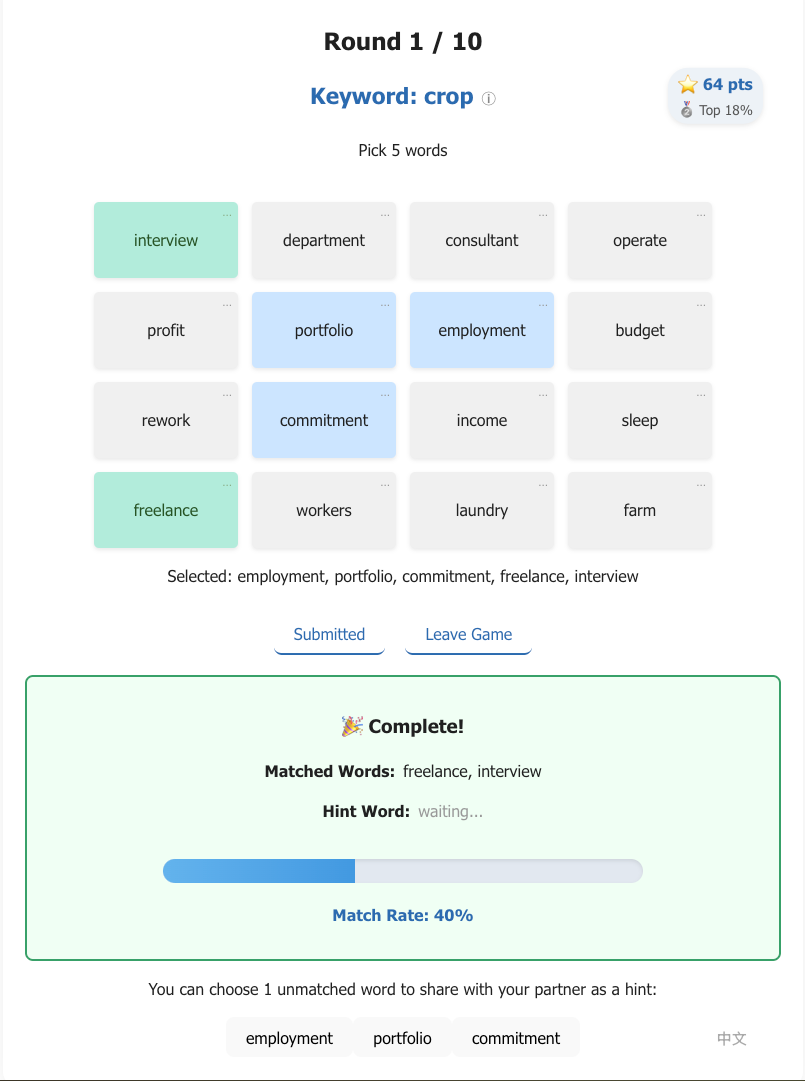}
  \caption{The word choice interface displays a grid of words from which the user selects based on the given clue, the round result section displays the match rate and highlight matched words between the players.}
  \label{fig:word_choice}
\end{figure}

\begin{figure}[H]
  \centering
  \includegraphics[width=0.8\linewidth]{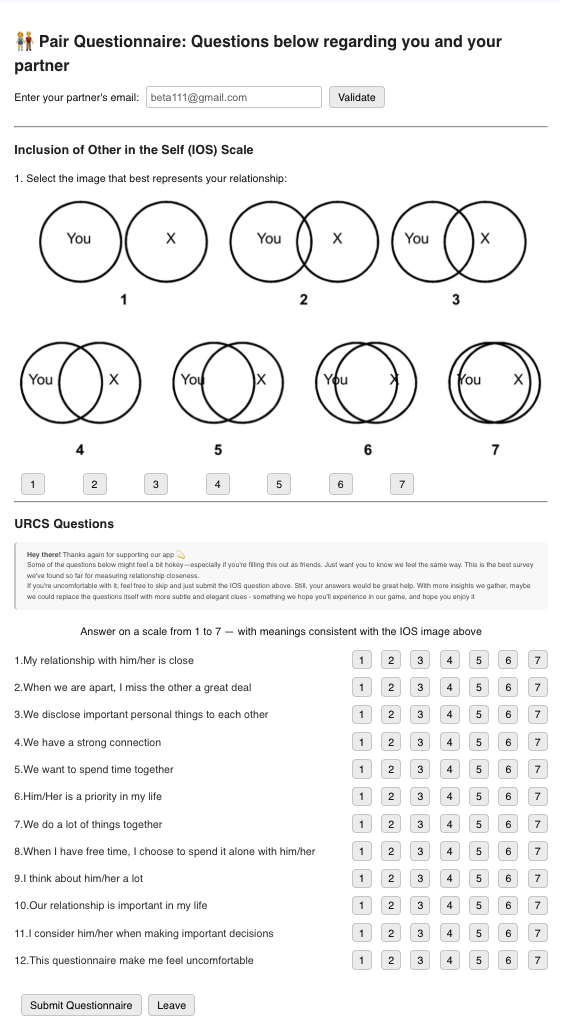}
  \caption{This screen shows how players complete pair questionnaire to provide compatibility labels.}
  \label{fig:data_collection}
\end{figure}

\begin{figure}[H]
  \centering
  \includegraphics[width=0.8\linewidth]{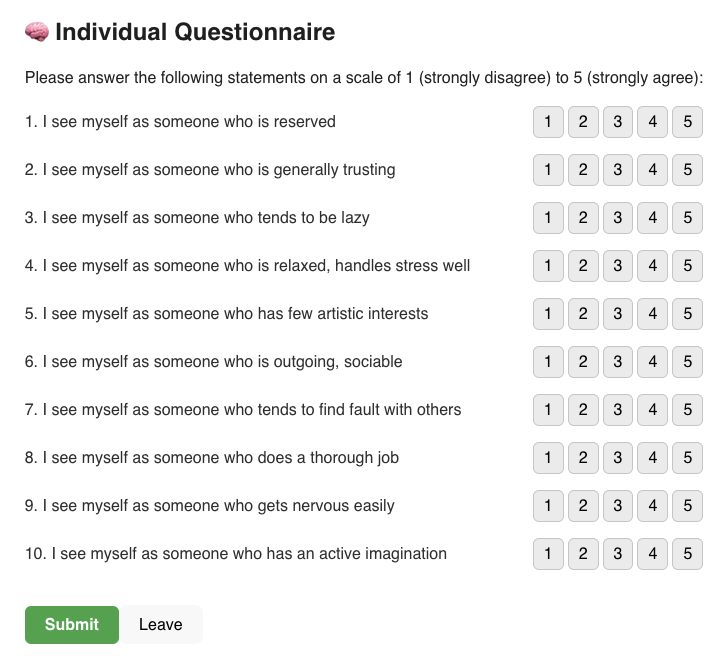}
  \caption{This screen shows how players complete individual questionnaire to provide personality traits label.}
  \label{fig:data_collection2}
\end{figure}

\bibliographystyle{ACM-Reference-Format}
\bibliography{sample-base}
\end{document}